\documentclass[a4paper,11pt]{article}
\pdfoutput=1 % if your are submitting a pdflatex (i.e. if you have
\usepackage{jheppub} % for details on the use of the package, please
\usepackage{color}
\usepackage{graphicx}
\usepackage{amsmath}
\usepackage{amssymb}
\usepackage[normalem]{ulem}
\definecolor{darkred}{rgb}{0.6,0,0}

\renewcommand\thesection{\arabic{section}}

\definecolor{linkcolor}{rgb}{0,0,0.5}

\def\gsim{\raise0.3ex\hbox{$\;>$\kern-0.75em\raise-1.1ex\hbox{$\sim\;$}}}
\def\lsim{\raise0.3ex\hbox{$\;<$\kern-0.75em\raise-1.1ex\hbox{$\sim\;$}}}

\def\beqn#1{\begin{equation}\label{#1}}
\def\eeqn{\end{equation}}

\def\beqa#1{\begin{eqnarray}\label{#1}}
\def\eeqa{\end{eqnarray}}

\def\Z2{$\mathcal{Z_2}$}

\def\vev#1{\left\langle #1\right\rangle}

\newcommand {\ignore}[1]{}

\newcommand{\AddrAHEP}{%
	AHEP Group, Institut de F\'{i}sica Corpuscular --
	CSIC/Universitat de Val\`{e}ncia, Parc Cient\'ific de Paterna.\\
	C/ Catedr\'atico Jos\'e Beltr\'an, 2 E-46980 Paterna (Valencia) - SPAIN}
\begin{document}
\title{On the high-scale instanton interference effect: axion models without domain wall problem}

\author{Mario Reig}\emailAdd{mario.reig@ific.uv.es}
\affiliation{\AddrAHEP}

%%%%%%%%%%%%%%%%%%%%%%%%%%%%%%%%%%%%%%%%%%%%%%%%%%%%%%%%%

%\begin{textblock*}{5cm}(11cm,-8.2cm)
%\end{textblock*}

%%%%%%%%%%%%%%%%%%%%%%%%%%%%%%%%%%%%%%%%%%%%%%%%%%
\abstract{We show that a new chiral, confining interaction can be used to break Peccei-Quinn symmetry dynamically and solve the domain wall problem, simultaneously. The resulting theory is an invisible QCD axion model without domain walls. No dangerous heavy relics appear.}

\maketitle
\flushbottom

%%%%%%%%%%%%%%%%%%%%%%%%%%%%%%%%%%%%%%%%%%%%%%%%%%%%%%%%%

\section{Introduction}
\label{sec:intro}
%%%%%%%%%%%%%%%%%%%%%%%%%%%%%%%%%%%%%%%%%%%%%%%%%%%%%%%%%

The appearance of topological defects during spontaneous breaking of symmetries constitutes a clear and profound connection between particle physics and cosmology \cite{Vilenkin:1984ib}. As the Universe cools down several phase transition take place and, depending on the homotopy groups of the manifold of degenerate vacua, stable topological defects may form \cite{Kibble:1976sj}.

In particular, the cosmic domain wall problem \cite{Zeldovich:1974uw} is a well-known potential issue of axion models \cite{Sikivie:1982qv,Vilenkin:1982ks}. Recently, it has been pointed out that Majoron models can also suffer from domain walls \cite{Lazarides:2018aev}. To solve such a long-standing problem, several mechanisms have been proposed. Being a cosmological-particle physics issue, it is not surprising that one can tackle it from both, cosmology and particle physics sides. A couple of well known solutions are: cosmic inflation and the Lazarides-Shafi mechanism \cite{Lazarides:1982tw}. In the first one the dangerous walls are pushed beyond the horizon, being a clear example of a cosmological solution. In the second case, one associates the spontaneously broken discrete symmetry to a gauge symmetry. This removes the physical degeneracy among the different vacua, which become gauge equivalent. Another interesting solution that has been recently suggested \cite{Sato:2018nqy} implements the Witten effect to solve the domain wall problem. More exotic post-inflationary solutions involve primordial black holes to perforate the walls, change their topology and destroy them \cite{Stojkovic:2005zh}.

As noted by Holdom \cite{Holdom:1982ew}, the cosmic domain wall problem seems to be associated to the breaking of symmetries by scalars. One can imagine that the degeneracy of the associated vacua disappears for theories where the breaking of Peccei-Quinn (PQ) symmetry is dynamically triggered by new confining forces. In addition, in usual invisible axion models \cite{Zhitnitsky:1980tq,Dine:1981rt,Kim:1979if,Shifman:1979if}, the introduction of a SM singlet condensate $\vev{\sigma}$ breaking PQ symmetry at high energies generates a fine-tuning problem. The bare mass terms of the Higgs isodoublets, $\mu_{u,d}^2|H_{u,d}|^2$, have to cancel almost perfectly (with EW scale precision) quartic couplings of the type $\lambda|\sigma|^2|H_{u,d}|^2$. This is nothing but the standard, well-known hierarchy problem of axion models. We conjecture that the solution of both problems is intimately related. The reason is that dynamical breaking of symmetries by fermion condensates usually brings associated instanton effects, which explicity break anomalous symmetries.  

Dynamical breaking of PQ symmetry has a long history  in the context of composite axion models \cite{Kim:1984pt} and has gained interest recently \cite{Yamada:2015waa,Wilczek:2016gzx,Redi:2016esr,Gaillard:2018xgk,Lee:2018yak,Anastasopoulos:2018uyu,Gavela:2018paw}. New strongly coupled and confining interactions have been also reconsidered recently to rise the QCD axion mass \cite{Rubakov:1997vp,Hook:2014cda,Fukuda:2015ana,Gherghetta:2016fhp,Dimopoulos:2016lvn,Agrawal:2017evu,Agrawal:2017ksf}.

In this work we build a minimal model where the breaking of PQ symmetry arises dynamically at a high scale thanks to a new chiral confining force. In addition, we show how the associated instantons implement the \textit{instanton interference effect} (IIE), solving the domain wall problem. A similar construction has been explored by Barr and Kim \cite{Barr:2014vva}. In this reference it was suggested that new confining interactions can solve the domain wall problem. However, despite it avoids the domain wall problem with a $N_{DW}=1$ scenario, it does illustrate the appearance of a phenomenological and cosmological problem, namely the overclosing of the Universe by heavy stable relics. This issue seems almost unavoidable in the context of confining interactions, since they usually bring associated conserved quantum numbers. Baryon number in the Standard Model is the most clear example.

If the lightest of these unconfined bound states, which we will call \textit{hyperbaryons}, is stable it might overclose the Universe depending on its mass. Following Griest and Kamionkowski \cite{Griest:1989wd}, this limit is given by:
\begin{equation}
\Lambda_{HC}\leq 240\text{ TeV} \,.
\end{equation}
Dangerous heavy relics can be diluted after a period of cosmic inflation. However, in such a case one might ask why we do not also use inflation to avoid the domain wall problem. We will show below which are the basic ingredients to achieve a phenomenologically successful solution to all these problems.

%%%%%%%%%%%%%%%%%%%%%%%%%%%%%%%%%%%%%%%%%%%%%%%%

\section{The model}
Let us consider a model based on the symmetry:
\begin{equation}
G\times U(1)_{PQ}\times SM\,,
\end{equation}
with the $SM$ factor being the local symmetry of the Standard Model, $SU(3)_C\times SU(2)_L\times U(1)_Y$, and $G$ a new confining gauge group. $U(1)_{PQ}$ is the global, anomalous PQ symmetry. All SM particles are $G$ singlets. For simplicity, we assume that the PQ charges for SM fermions are given by fermion chirality, this is +1 for left-handed (LH) fermions and  $-1$ for right-handed (RH) fermions\footnote{This particular choice is compatible with an underlying $SO(10)$ GUT \cite{Ernst:2018bib}.}. This assignment is compatible with the original PQWW and DFSZ axion model \cite{Peccei:1977hh,Wilczek:1977pj,Weinberg:1977ma}. 
In the $G$ sector, we assume one of the fermions has PQ charge $+1$ while the other has no PQ charge. The reason for this will become clear later. Therefore, the fermion content of the model is given by:
\begin{equation}\label{particles}
\begin{split}
q^i_L\sim (1,3,2,1/6)&_{1}\,,\,\,l^i_L\sim (1,1,2,-1/2)_{1}\,,\\
 u^i_R\sim (1,3,1,2/3)&_{-1}\,,\,\,d^i_R\sim (1,3,1,-1/3)_{-1}\,,\\
 e^i_R\sim (&1,1,1,-1)_{-1}\,,\\
  \psi_1\sim (R,1,1,0)&_{1}\,,\,\,\psi_2\sim (R,1,1,0)_{0}\,,
\end{split}
\end{equation}
where the first quantum number stands for $G$ and the subscript is the PQ charge. 
The distinction between $\psi_1$ and $\psi_2$ requires the coupling to different scalar fields.

It can be seen that  $U(1)_{PQ}$ symmetry is anomalous under both, QCD and $G$. This is a reasonable assumption, since it seems rather artificial to protect an anomalous symmetry from anomalies of another gauge group. The first question that arises is which kind of groups are appropriate for $G$. Many possibilities emerge. However, one has to deal with the limitations coming from the $G$ triangle anomaly. As we will see below $G=SO(4N+2)$, for $N\geq 2$, suggest themselves as the most natural choice. Strikingly enough, they admit complex, chiral  representations and are anomaly free. The use of spinor representations will also be important to solve the heavy relic problem. From now on, we will assume the $G$ gauge group is given by the well known group:
\begin{equation}
G=SO(10)_{HC}\,,
\end{equation}
and we will refer to it as $\textit{hypercolor}$ (HC). This interaction becomes strongly coupled and confining at the HC scale, $\Lambda_{HC}$.

The HC fermions previously mentioned in Eq.(\ref{particles}) are a couple of $SO(10)_{HC}$ spinors, $\psi_1\sim (16,1,1,0)_1$ and $\psi_2\sim (16,1,1,0)_0$. The scalars required to distinguish them are $SO(10)_{HC}$ vectors \begin{equation}\label{SO(10)scalars}\begin{split}
&H_{1}\sim (10,1,1,0)_{2}\,,\\&
H_{2}\sim (10,1,1,0)_{0}\,.
\end{split}\end{equation}
Therefore, the Yukawa lagrangian of the HC sector is given by:
\begin{equation}
\mathcal{L}_{HC}=y_1\bar{\psi_1}^cH_1^\dagger\psi_1+y_2\bar{\psi_2}^cH_2^\dagger\psi_2\,.
\end{equation}
Notice that due to PQ charge assignement $\psi_1$ and $\psi_2$ do not mix. While the scalar $H_1$ develops a non-zero vacuum expectation value (vev) below the HC scale 
\begin{equation}
\vev{H_1}<\Lambda_{HC}\,,
\end{equation}
we assume that $H_2$ has an inverted phase transition as in \cite{Barr:2014vva}. This inverted phase transition is characterized by a non-zero vev, $\vev{H_2}\neq 0$, above a critical cosmic termperature $T\geq T_c$. Below this temperature, $T\leq T_c$, the vev vanishes $\vev{H_2}=0$ and, since gauge symmetry does not allow a bare mass term for SO(10) spinors, $\psi_2$ becomes exactly massless. It was shown by Weinberg that this kind of phase transitions can exist \cite{Weinberg:1974hy}. We also require that the HC confinement scale is larger than the critical temperature, $\Lambda_{HC}\geq T_c\gg \Lambda_{QCD}$ so that PQ and the chiral symmetry of $\psi_2$ do not coexist as classical symmetries of the lagrangian\footnote{Recently a low-scale version with $\Lambda_{HC}\ll\Lambda_{QCD}$ has also been proposed \cite{Caputo:2019wsd}.}. 
%As we will see later, $\psi_1$ also gets a non-zero mass from HC dynamics.

As in the $SU(N)$ family, $SO(N)$ gauge theories have non-trivial vacuum for $N\geq 3$. This can be seen from their non-trivial homotopy group, e.g. $\pi_3(SO(N))=\mathcal{Z}$ (with the exception $\pi_3(SO(4))=(\mathcal{Z})^{\times 2}$). There are two different $\theta$-terms, one for QCD and one for HC. Since we are imposing a unique PQ symmetry it might seem that we are not solving the strong CP problem, as one usually needs the same number of anomalous symmetries than $\theta$-terms or confining interactions. However, being a chiral representation, the spinor does not allow a bare mass term. 
As we have seen, this implies that $\psi_2$ is massless after the inverted phase transition (below $T_c$). It is well known that the topological susceptibility vanishes when there is a fermion with zero mass \cite{Schafer:1996wv}. Since the topological susceptibility is the second derivative of the vacuum energy respect to $\theta_{HC}$, the theory is $\theta_{HC}$ independent below $T_c$. This fact renders $\theta_{HC}$ unphysical while $\theta_{QCD}$ is driven dynamically to zero by the standard PQ mechanism. This situation resembles the one proposed by Barr and Kim in their work \cite{Barr:2014vva}.

%
%%%%%%%%%%%%%%%%%%%%%%%%%%%%%%%%%%%%%%%%%%%%%%%%%%%%%%%%%

\section{The instanton interference effect (IIE)}
\begin{figure*}[t]
	\centering
	\includegraphics[width=0.39\textwidth]{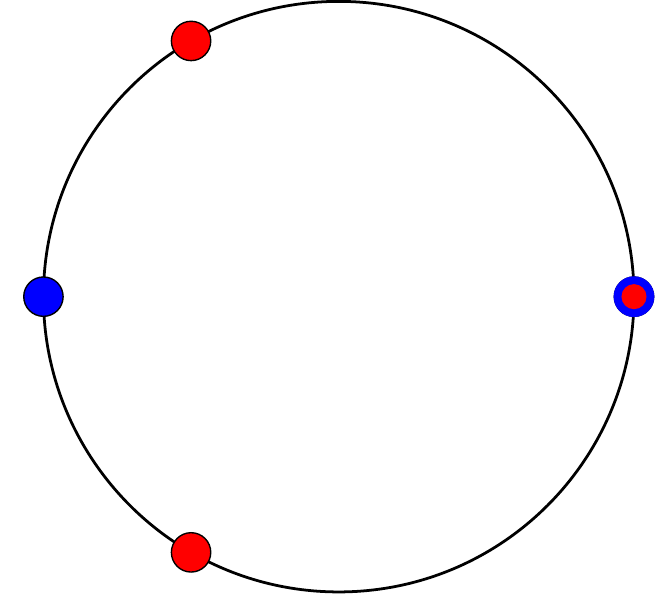}
	\caption{The IIE pitorically: a given U(1) is broken, by instantons, to a $Z_N$ and a $Z_M$ discrete symmetries. If $N$ and $M$ are co-prime numbers there is no common subgroup and the $U(1)$ is completely broken. This example illustrates the mechanism for $N=2$ and $M=3$. It can be easily seen that the blue and red dots only coincide in one point (mod $2\pi$).}
	\label{instanton_interference}
\end{figure*}

Since we are extending the QCD axion model with a new confining interaction there are two potential sources of explicit PQ symmetry breaking. In general, QCD and HC instantons will break $U(1)_{PQ}$ symmetry explicitly, generating two independent cosine-like potentials for the axion field: $V_{QCD}(a)$ and $V_{HC}(a)$ with periodicities $2\pi/N_{QCD}$ and $2\pi/N_{HC}$, respectively. Therefore, the breaking of $U(1)_{PQ}$ is done in the direction of $Z_{N_{HC}}$ and $Z_{N_{QCD}}$ discrete subgroups, with $N_{HC}$ and $N_{QCD}$ the anomaly coefficients. This may lead to an interesting situation where the explicit breaking is not in the direction of the same subgroup. The residual discrete symmetry unbroken by the combination of non-perturbative effects will be the common subgroup of both, $Z_{N_{HC}}$ and $Z_{N_{QCD}}$. This is what we call \textit{instanton interference} and can be pictorially visualized in figure \ref{instanton_interference}. If the anomaly coefficients $N_{HC}$ and $N_{QCD}$ are co-prime numbers, the \textit{instanton interference} completely solves the domain wall problem, since  $Z_{N_{HC}}$ and $Z_{N_{QCD}}$ have no common subgroup. The anomaly coefficients are given by 
\begin{equation}\label{anomaly}
N=2\sum_R q_Rt_R\,,
\end{equation}
with $t_R$ the Dynkin index of the representation $R$, defined in terms of the group generators as $\text{Tr}[T^aT^b]=t_R\delta^{ab}$, and $q_R$ the PQ charge of $R$. For $SU(N)$ groups the indices are $t_N=1/2$ and $t_{adj}=N$ for the fundamental and adjoint representation, respectively. On the other hand, for $SO(2n)$ the indices are given by $t_{2n}=1$ and $t_{spinor}=2^{n-4}$, for the fundamental and spinor representations.

There is, however, a subtlety related to the high-scale IIE. If one wants to preserve the PQ solution to the strong CP problem the HC interaction must be turned off below a certain critical temperature, $T_c\gg\Lambda_{QCD}$, as explained above. In other words, there is no epoch when both HC and QCD potentials are turned on. This implies that the high-scale IIE works only when $N_{HC}=1$\footnote{Note that this is not the case for the low-scale version of the IIE \cite{Caputo:2019wsd} where $N_{HC}$ can be different from 1 as long as it is relatively prime to $N_{QCD}$. This is because for the low-scale version, the HC and QCD axion potentials are both turned on at low energies. See \cite{Caputo:2019wsd} for details.}. In this case, the axion string-wall system decays soon after it is formed, at $T\sim\Lambda_{HC}$, and the axion field takes the same value everywhere. Consequently, when QCD instantons turn on at much smaller temperatures, $T\sim\Lambda_{QCD}$, no axionic domain wall can form because the minimum of the axion field is the same even in causally disconnected regions thanks to the interference effect of the HC instantons.

In the case of the HC sector of the previous section, a straightforward calculation give us $N=4$. However, one realizes that these vacua are actually related by a gauge transformation (this is because the HC gauge invariant order parameter has PQ charge equal 4) leading to an effective $N_{HC}=1$, as the high-scale IIE requires.
%%%%%%%%%%%%%%%%%%%%%%%%%%%%%%%%%%%%%%%%%%%%%%%
\subsection{Scalar potential and $N_{QCD}$}
The scalar potential of the model described above is relatively simple. Since the model is DFSZ-like in the SM sector, the part of the scalar potential corresponding to the Higgs doublets coupling to SM fermions is given by\footnote{We do not consider, at this point, quartic and bare mass terms involving $H_1$ and $H_2$ since they are not relevant for the determination of $Z_{N_{QCD}}$.}:
\begin{equation}\begin{split}
&V(\phi_u,\phi_d)=\mu^2_u|\phi_u|^2+\mu_d^2|\phi_d|^2+\lambda_u|\phi_u|^4+\lambda_d|\phi_d|^4\\&+\lambda_1|\phi_u|^2|\phi_d|^2+\lambda_2(\phi_u\phi_d)(\phi_u\phi_d)^\dagger+h.c\,,
\end{split}\end{equation}
where
\begin{equation}
\phi_u\sim (1,2,-1/2)_2\,,\,\,\phi_d\sim(1,2,1/2)_2\,,
\end{equation}
couple to up-type and down-type fermions, respectively. The degeneracy of the vacuum is determined by the gauge invariant order parameters. This potential, in combination with the PQ charges of (\ref{particles}) and the anomaly coefficient computed as dictated by Eq.(\ref{anomaly}), reveals a $Z_{N_{QCD}}=Z_3$ symmetry among the different vacua. Therefore, we have a situation with $N_{HC}=1$, $N_{QCD}=3$ and the IIE  solves the domain wall problem as explained above.

To avoid the emergence of an extra, unwanted U(1) symmetry it is crucial to connect the HC sector, in particular the $\psi_1$ and $H_1$ fields, to the SM. This is done by the quartic term 
\begin{equation}\label{portal}
\Delta V=\lambda_{mix}\,(\phi_u\phi_d)^\dagger H_1H_1+h.c.
\end{equation}	
Then the PQ symmetry is connected to the HC sector and the IIE operates. This term is also important to connect the spontaneous breaking of PQ symmetry to the HC scale and not to the EW scale. 

A last comment regarding the scalar sector is that it might seem one can redefine PQ symmetry by taking into account an apparent chiral $U(1)_{\psi_2}$ symmetry associated to $\psi_2$. Such a redefinition would be broken only by QCD instantons in a similar way to the standard dynamical axion \cite{Kim:1984pt}. This is however not necessarily the case. One can easily imagine that due to terms like $\mu^2 H_2H_2$ or $\lambda |H_1|^2H_2H_2$ (allowed by gauge invariance and breaking explicitly $U(1)_{\psi_2}$ in the potential) this redefinition of PQ symmetry is not allowed. In this case, PQ symmetry is necessarily anomalous under both QCD and HC and the IIE works to solve the domain wall problem. Note that even in this case there is still a $Z_4$ symmetry, with $\psi_2$ and $H_2$ transforming as 
\begin{equation}
\psi_2\rightarrow e^{i\pi/2}\psi_2\,,\,\,\,H_2\rightarrow -H_2\,,
\end{equation}
that prevents the $\psi_2$ fermion of getting a mass below $T_c$.

%%%%%%%%%%%%%%%%%%%%%%%%%%%%%%%%%%%%%%%%%%%%%%%%%%%%

\subsection{The $\eta\prime_{HC}$ and the axion}
As we have seen before the $\theta_{HC}$ term is not physical due to the massless HC fermion, $\psi_2$. This makes the axion potential coming from HC instantons flat below $T_c$.
In the dual description, after confinement, this $\theta_{HC}$-term is relaxed dynamically by the $\eta^\prime_{HC}$, which acquires a mass of the order $\Lambda_{HC}$ and decouples from the low-energy theory. With the axion potential coming from HC instantons, $V_{HC}(a)$, turned off below $T_c$, only QCD effects generate a mass for the axion giving the usual QCD prediction \cite{diCortona:2015ldu}:
\begin{equation}
m_{a}^2\sim \frac{m_um_d}{(m_u+m_d)^2}\frac{f_\pi^2m_\pi^2}{f_a^2}\,.
\end{equation} 
As we will see below, $f_a$ is also identified with $\Lambda_{HC}$, since the condensate $\vev{\psi_1\psi_1}\sim \Lambda_{HC}^3$ breaks PQ symmetry spontaneously. The HC scale, $\Lambda_{HC}\sim f_a$, is presumably very large making the QCD axion naturally invisible.
\section{Avoiding heavy stable relics}
New confining interactions can have a non-trivial cosmological impact. For $SU(N)$ gauge groups, a global $U(1)$ symmetry analogous to baryon number protects the stability of the lightest hyperbaryon. As an example, a $SU(N)$ theory with a fermion in the fundamental $N$ representation has a stable hyperbaryon composed by $N$ fermions. Analogously, $SO(N)$ groups feature a $Z_2$ conserved quantum number. This $Z_2$ symmetry counts the number of indices for the fundamental representation and can stabilize the lightest bound state. These group theoretic properties have been used to stabilize dark matter \cite{Antipin:2015xia}. Another example of exotic matter stabilized by the conserved $Z_2$ symmetry of an $SO(N)$ confining interaction are hyperbaryons in the context of Comprehensive Unification \cite{Reig:2017nrz}.

It was shown long ago by Griest and Kamionkowski that a stable particle that was in thermal equilibrium may overclose the Universe if it is very heavy \cite{Griest:1989wd}. By using partial wave unitarity of the S-matrix, one can estimate the relic density today as:
	\begin{equation}
	\Omega_\chi h^2\approx 10^{-5}\left(M_\chi /\text{TeV}\right)^2\,.
	\end{equation}
Such a stable relic $\chi$ would produce an unacceptable DM relic density $\Omega_\chi h^2>\Omega_{DM}h^2$ or even  $\Omega_\chi\geq 1$ if it is heavier than $\mathcal{O}(100)$ TeV.	  

One can naively think that because $SO(10)_{HC}$ is confining, there will be stable hyperbaryons. However, this is not the case. When the interaction becomes strongly coupled, HC dynamics will form the condensate:
\begin{equation}\label{condensate}
\vev{\psi_1\psi_1}\sim \Lambda_{HC}^3\,.
\end{equation}
As mentioned before, this condensate breaks $U(1)_{PQ}$ spontaneously. Additionally, the condensate has also non-trivial implications in the dynamics of $SO(10)_{HC}$ itself. Here we briefly describe the situation. In the single gauge boson exchange approximation the potential between two $\psi$ is given by:
\begin{equation}
V\sim \frac{g_{HC}^2}{2r}\left[C_c- C_{16} - C_{16}\right]\,,
\end{equation}
with $C_{16}$ and $C_c$ the Casimir invariants of the spinor and the possible representations of the condensate. Interestingly this condensate does not contain an $SO(10)$ singlet since: 
\begin{equation}
16\times 16=10_{s}+120_{a}+126_{s}\,.
\end{equation}
Therefore, the condensate must break $SO(10)_{HC}$ gauge symmetry. This is done following the most attractive channel (MAC) rules \cite{Raby:1979my}. Since the only attractive channel is in the 10 representation direction, which contains a $SO(9)$ singlet, the symmetry breaking reads:
\begin{equation}
\begin{split}
SO(1&0)\rightarrow SO(9)\,,\\&
16\rightarrow 16\,.
\end{split}
\end{equation}
The strongly coupled $SO(9)$ interaction, again, rearrange the vacuum and form condensates $\vev{\psi_1\psi_1}$\footnote{The MAC rules state that the fermions involved in the condensate get a non-zero mass. Therefore we need to assume that only $\psi_1$ form condensates while $\psi_2$ remains massless.}. However, this condensate now contains an $SO(9)$ singlet, since:
\begin{equation}
16\times 16=1_s+9_s+36_a+84_a+126_s\,.
\end{equation}
The SSB chain of the confining interaction stops at this point and the $SO(9)$ remains unbroken. 
%Interestingly, this condensate introduces a linear term in the potential for $H_1$:
%\begin{equation}
%V=-y_1\vev{\psi_1\psi_1}H_{1}+\mu^2|H_1|^2+\lambda|H_1|^4\,.
%\end{equation}
%In the limit $\lambda\ll 1$, the condensate induces a non-zero vev for $H_1$, $\vev{H_1}=y_1\vev{\psi_1\psi_1}/\mu^2$, generating a perturbative mass for $\psi_1$ from HC dynamics.

We are now about to show why no dangerous heavy relics emerge in our framework. For $SO(N)$ with $N$ odd there are two conjugacy classes. The conjugacy classes of the spinor and singlet representations are $C=1$ and $C=0$, respectively. Since the product of representations decomposes into representations with the same class (mod 2), only products of an even number of spinors can give us $SO(9)$ singlets. Therefore, since $SO(9)$ supports a conserved $Z_2$ quantum number instead of $U(1)$, no stable $SO(9)$ singlet can appear. All the possible hyperbaryons, i.e. $SO(9)$ singlet bound states, are $Z_2$ singlets and decay.  

In more detail, since both $\psi_1$ and $\psi_2$ are lighter than the HC scale the lightest bound states are \textit{HC-mesons} composed by two SO(9) spinors. Note this differs from a pure Yang-Mills theory (or a theory with fermions much heavier than the confinement scale) where the lightest bound states are glueballs. HC-mesons are allowed to decay into SM degrees of freedom through HC anomaly diagrams (in a similar way to the QCD pion) and the interaction in Eq. \ref{portal} (or similar quartic terms), which connects the HC and SM sectors. Since HC-mesons have tipically masses of the order of the HC scale $\Lambda_{HC} $, their lifetime is expected to be extremely short.

%%%%%%%%%%%%%%%%%%%%%%%%%%%%%%%%%%%%%%%%%%%%%%%%%%%%%%%%%
\section{HC coupling running and confinement}
\begin{figure*}[t]
	\centering
	\includegraphics[width=0.59\textwidth]{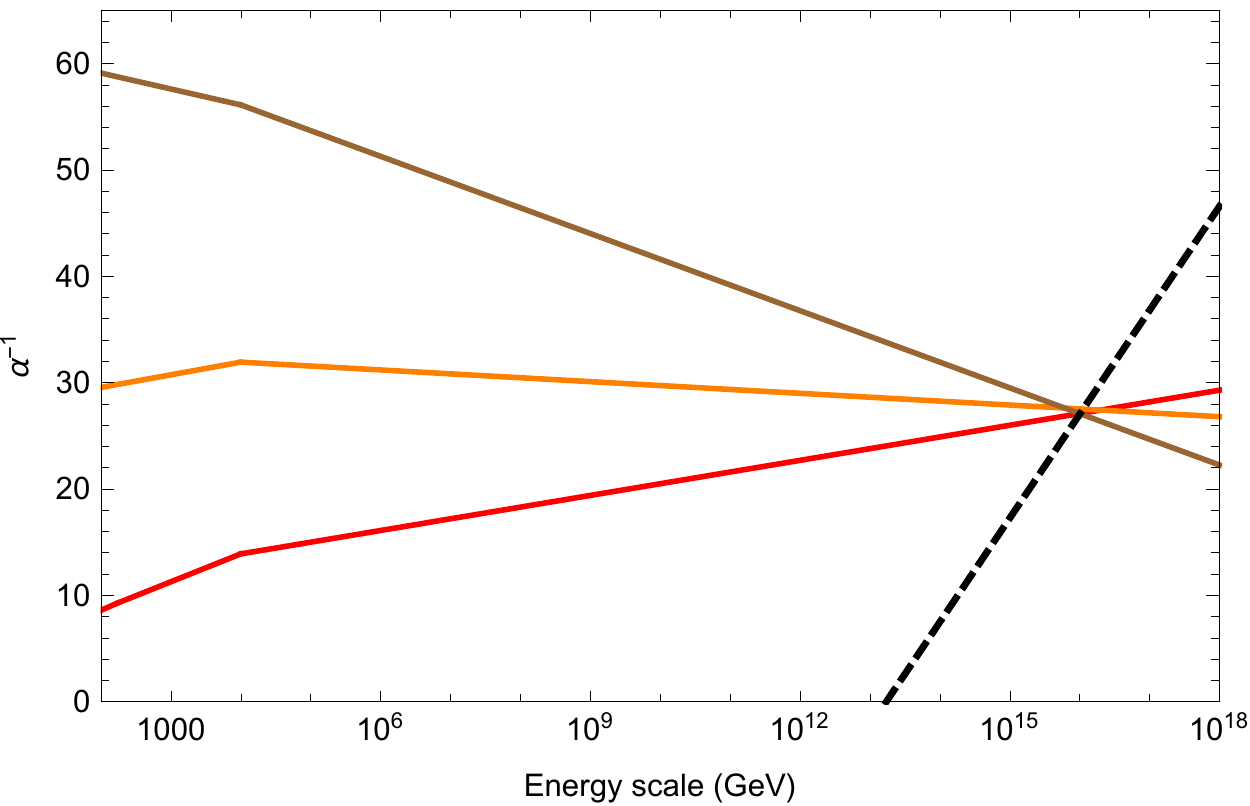}
	\caption{Running of gauge couplings in a supersymmetric scenario. HC gauge coupling corresponds to the dashed black line. See text.}
	\label{Plot}
\end{figure*}

We have assumed that the HC interaction becomes strongly coupled at high energies. For the sake of completeness, let us study an explicit example quantitatively. If some kind of new physics like supersymmetry or new scalars is responsible of the unification of gauge couplings at high energies, it is attractive to imagine that the HC interaction is also unified with the other SM interactions at around $M_{GUT}\sim 10^{16}$ GeV. At one loop and neglecting threshold corrections, the evolution of the couplings is governed by:
\begin{equation}
\alpha^{-1}(\mu)=\alpha^{-1}(M)+\frac{1}{6\pi}\ln\frac{M}{\mu}\left[-11\,C_2(G)+2T(R)\right]\,,
\end{equation}
with $C_2(SO(N))=N-2$, and $T(\text{spinor})=2^{N/2-4}$. With two HC spinors and no light scalars, this give us:
\begin{equation}
\alpha_{HC}^{-1}(\mu)=\alpha_{HC}^{-1}(M)-\frac{80}{6\pi}\ln\frac{M}{\mu}\,.
\end{equation} 
Taking a supersymmetric example, if $\alpha_{HC}(M_{GUT})=1/28$ and $M_{GUT}=10^{16}$ GeV, one can estimate the scale $\Lambda_{HC}$ at which the HC coupling becomes strong as:
\begin{equation}
\Lambda_{HC}\approx e^{-27\times 6\pi/80}\times 10^{16}\,.
\end{equation}
We obtain $\Lambda_{HC}\approx 1.7\times 10^{13}$ GeV, which lies close to the upper bound of the axion window (see Fig. \ref{Plot}). The HC scale can be lowered if $\alpha(M_{GUT})<1/28$ or if there is extra matter in the HC sector below the GUT scale, making the running of the coupling slower. Then, one can easily obtain a HC scale $\Lambda_{HC}$ inside the QCD axion window.

Obviously, the coincidence of all interaction strengths in one point of the $(\alpha^{-1},E)$ plane is meaningless unless there is an underlying unified gauge group in the UV. The reason is that if there is no unified group containing HC and the SM, keeping the coupling evolution together above $M_{GUT}$, the gauge couplings will separate as can be seen in Fig. \ref{Plot}. However, we believe that this example does illustrate the principles of model building.
%%%%%%%%%%%%%%%%%%%%%%%%%%%%%%%%%%%%%%%%%%%%%%%%
\section{Discussion}
%%%%%%%%%%%%%%%%%%%%%%%%%%%%%%%%%%%%%%%%%%%%%%%%%%%%%%%%%%
Before closing we comment on different aspects of the model that deserve mention:
\begin{itemize}
	\item In an hypothetical $SO(10)_{HC}\times U(1)_{PQ}\times SO(10)$ theory, the unbroken group by the instanton interference mechanism is a $Z_2$ symmetry that can be automatically associated to the center of both $SO(10)$ groups. Then, the Lazarides-Shafi mechanism \cite{Lazarides:1982tw} can be naturally implemented without adding extra fermions. 

	\item	It is attractive to imagine that some sort of interaction with the condensate $\vev{\psi\psi}$ can generate small neutrino masses. A plausible possibility is to use $\psi$ in a radiative mechanism, together with the appropriate scalars, in close analogy to the mechanism presented in \cite{Reig:2018ztc}. 
\end{itemize}
%%%%%%%%%%%%%%%%%%%%%%%%%%%%%%%%%%%%%%%%%%%%%
\section{Conclusions}
New chiral confining interactions with fermions in the spinor representation can simultaneously make the axion invisible and solve the domain wall problem. The instanton interference effect (IIE) has been described in detail. If PQ symmetry remains unbroken during inflation, the instanton interference mechanism suggests itself as a compelling possibility to avoid the domain wall problem, combining different sources of explicit breaking by instantons. Finally, spinor representations of chiral, anomaly free groups turn out to be the  crucial ingredient to explain the absence of heavy stable relics. This fact strongly suggest them as compelling candidates for the confining interaction of composite axion models.

\begin{acknowledgments}
I am especially grateful to R. Fonseca for helpful discussions about group theory and P. Agrawal for discussions and insightful comments during the early stages of this work. I am also grateful to M. Yamada and P. Quilez for enlightening discussions about a preliminary version of the manuscript. I would also like to thank Stockholm University and the organizers of the "Quantum Connections: Axions in Stockholm" workshop, where this work started, for hospitality. This work is supported by the Spanish grants FPA2017-85216-P (AEI/FEDER, UE), SEV-2014-0398 and PROMETEO/2018/165 (Generalitat  Valenciana). 
\end{acknowledgments}

%\bibliographystyle{utphys}
%\bibliography{newrefs,newrefs_axion} 

\providecommand{\href}[2]{#2}\begingroup\raggedright\endgroup

\end{document}